\documentclass[useAMS,usenatbib]{mn2e}

\usepackage {psfig}
\usepackage{amssymb}
\usepackage {graphicx}
\usepackage {subfigure}
\usepackage {epsfig}
\usepackage{mathrsfs}
\usepackage[english]{babel}
\usepackage{amsfonts}
\usepackage{fontenc}
\usepackage{textcomp}
\usepackage{microtype}
\usepackage[pdftex]{lscape}

\title[Relaxation of $N$-body systems with additive $r^{-\alpha}$ interparticle forces]{Relaxation of N-body systems with additive $r^{-\alpha}$ interparticle forces}
\author[P. F. Di Cintio, L. Ciotti and C. Nipoti]{PierFrancesco Di Cintio$^{1}$\thanks{E-mail:
pdicint@pks.mpg.de}, Luca Ciotti
$^{2}$, and Carlo Nipoti$^{2}$\\
$^{1}$Max Planck Institute for the Physics of Complex Systems, N\"othnitzer Stra\ss e 38, D-01187 Dresden Germany\\
$^{2}$Department of Physics and Astronomy, Bologna University, viale Berti-Pichat 6/2, I-40127 Bologna Italy}
\begin{document}

\date{Submitted 2013, January 22; Resubmitted 2013, March 1}

\pagerange{\pageref{firstpage}--\pageref{lastpage}} \pubyear{0000}

\maketitle

\label{firstpage}

\begin{abstract}
In Newtonian gravity the final states of cold dissipationless collapses are characterized by several structural and dynamical properties remarkably similar to those of observed elliptical galaxies. Are these properties a peculiarity of the Newtonian force or a more general feature of long-range forces? We study this problem by means of $N-$body simulations of dissipationless collapse of systems of particles interacting via additive $r^{-\alpha}$ forces. We find that most of the results holding in Newtonian gravity are also valid for $\alpha\neq2$. In particular the end products are triaxial and never flatter than an E7 system, their surface density profiles are well described by the S\'ersic law, the global density slope-anisotropy inequality is obeyed, the differential energy distribution is an exponential over a large range of energies (for $\alpha\geq1$), and the pseudo phase-space density is a power law of radius. In addition, we show that the process of virialization takes longer (in units of the system's dynamical time) for decreasing values of $\alpha$, and becomes infinite for $\alpha=-1$ (the harmonic oscillator). This is in agreement with the results of deep-MOND collapses (qualitatively corresponding to $\alpha=1$) and it is due to the fact the force becomes more and more similar to the $\alpha=-1$ case, where as well known no relaxation can happen and the system oscillates forever.
\end{abstract}


\section{Introduction}
One of the most striking properties of elliptical galaxies is the remarkable quasi-homology of their surface brightness profiles, described by the so called S\'ersic model, a generalization of the de Vaucouleurs $R^{1/4}$ model (see e.g. \citealt{caon93}, \citealt{andre95}, \citealt{courteau96}, \citealt{graham97}, \citealt{prugniel97}, \citealt{graham98}, \citealt{trujillo01}, \citealt{bert02}; see also \citealt{ciotti09} and references therein). Albeit minor (but important) departures from the S\'ersic model are common, overall the profiles on large scale are very well represented by the S\'ersic law. What is the origin of such regularity? $N$-body numerical simulations revealed that cold dissipationless and collisionless collapses lead to virialized end-states described almost perfectly by $R^{1/4}$ profiles (e.g. \citealt{vanAlbada82}, \citealt{LMS91}). More recently it has been shown that collapses in pre-existing dark matter halos are also well described by the S\'ersic profile, with a wide range of values of the S\'ersic index (Nipoti et al. 2006ab, hereafter N06ab). In addition, it is also known that the S\'ersic family is characterized by an exponential differential energy distribution, over a large range of accessible energies (e.g. \citealt{binney82}; \citealt{ciotti91}). These results can be understood in terms of the physics of violent relaxation in collisionless collapses (e.g. \citealt{LB67}, \citealt{bert84}, \citealt{bert03}, \citealt{bt05}, \citealt{bert05}). Finally, it has been proved analytically that in Newtonian gravity a large class of spherically symmetric equilibrium systems are characterized by the so-called Global Density Slope-Anisotropy Inequality (hereafter GDSAI, see Ciotti \& Morganti 2010ab; \citealt{vanhese11}, \citealt{an12}; see also \citealt{anevans}), a constraint between their anisotropy and density profiles. Numerical simulations suggest that the GDSAI may be a much more general result, holding true also for the final states of dissipationless collapses (see e.g. \citealt{hanmoo}).\\

Due to the relevance of these results for the understanding of the process of collisionless relaxation, a natural question arises about their apparent universality. In particular, are the S\'ersic law, the associated differential exponential energy distribution and the GDSAI peculiar features of Newtonian gravity or are they more general properties of the virialized final states of $N$-body collapses in which the particles interact with long-range forces? Preliminary results seem to support the second possibility. For example, it is known that the end-products of cold collapses in Modified Newtonian Dynamics (MOND, \citealt{BM84}) also produce final systems described remarkably well by the S\'ersic law (\citealt{NLC07}, hereafter N07a; and \citealt{CNL07}). However, the $N-$body MOND simulations have also shown that the oscillations leading to relaxation last more (in units of the dynamical time of the system) than in the equivalent Newtonian system (see e.g. N07a; \citealt{NLC07bis}). Moreover \cite{bar12} found indications that the GDSAI may be a common property of MONDian virialized systems. We recall that the force law in the MOND weak field limit (or deep-MOND, hereafter dMOND) regime is qualitatively similar to a force decreasing with distance as $1/r$.

Additional indications in this direction come from the 
preliminary analysis of Di Cintio (2009, hereafter DC09) and  Di Cintio \& Ciotti (2011, hereafter DCC11), who investigated the relaxation
of a system of spherical shells interacting via a long-range $r^{-\alpha}$ force
law, in analogy with similar studies performed in Newtonian gravity (\citealt{hen64}, \citealt{tak97}, \citealt{YM2000}) and MOND (Sanders 1998, 2008; Malekjani et al. 2009, 2012). In DCC11 we focused on the time evolution of the virial ratio and of the differential energy distribution. Among the main results, we confirmed the expectation that the process of relaxation, independently of the value of $\alpha$ (with the exception of $\alpha=-1$), consists in a first phase of violent collapse, followed by a longer, gentle phase of dynamical mixing. Remarkably, small values of $\alpha$ correspond to larger and long-lasting virial oscillations, confirming the dMOND results (N07a). However, the final states of shell systems are only poorly described by an exponential differential energy distribution. This is not surprising since the enforced spherical symmetry reduces the number of degrees of freedom available for energy exchanges during virialization.\\

Prompted by these preliminary results, here we explore further the problem following the collapse and virialization of fully three-dimensional $N$-body systems of particles interacting with radial forces proportional to a power-law of the mutual separation, $r^{-\alpha}$. This approach is not new, and here we recall the study of quasistationary states (\citealt{gabri10}, \citealt{gabri12}) and of Yukawa-like gravity (\citealt{moff96} and references therein; see also \citealt{BA12}). For the simulations we developed a direct $N$-body code, exploiting the force additivity. Note that for general forces, more sophisticated methods, based on the expansion in orthogonal functions of the potential, are not available, as the analogue of the Poisson equation does not exist. In our case, the considered forces even though additive, are described by {\it non-local} operators, i.e. the density at a given point can not be expressed as a simple differential operator of the potential at that point. In fact, the potentials associated with $r^{-\alpha}$ forces are the well known {\it Riesz potentials}, and the analogue of the Poisson equation involves the so called fractional Laplacian (e.g. \citealt{stein70}). Remarkably, for this specific cases, the potential can be expressed in terms of Gegenbauer polynomials in turn expressible with an addition theorem on spherical harmonics, so that in principle a multipole based Treecode can be realized (see \citealt{srinirasan}). Note that MOND is a non-linear but {\it local} theory.\\

The paper is organized as follows. In Section 2 we introduce the most important integral identities that will be used to study the results of the simulations, while in Section 3 the numerical code and the set-up of the initial conditions are presented. In Section 4 the virialization process and the structure and the dynamical properties of the virialized final states are presented and discussed as a function of $\alpha$. The main results are finally summarized in Section 5.
\section[]{Setting the stage}
We integrate numerically the equations of motion for an initially spherical system consisting of $N$ particles of identical mass $m$, mutually interacting with central long-range forces, obeying the superposition principle.
In particular the acceleration at  $\mathbf{r}_i$ due to a particle of mass $m_j$ at $\mathbf{r}_j$ is
\begin{equation}\label{eq1}
{\bf a}_{ji}=-Gm_j\times\frac{\mathbf{r}_i-\mathbf{r}_j}{||\mathbf{r}_i-\mathbf{r}_j||^{\alpha+1}},
\end{equation}
where $||~||$ is the standard Euclidean norm and $G$, is the force constant. The associated potential is
\begin{eqnarray}\label{eqp}
 \phi_{ji}=G m_j\times \cases{
 \displaystyle{
 {||\mathbf{r}_i-\mathbf{r}_j||^{1-\alpha }\over1-\alpha}, \quad \alpha\neq1,}\cr
  \ln||\mathbf{r}_i-\mathbf{r}_j||, \quad \alpha=1,
}
\end{eqnarray}
with $\mathbf{a}_{ji}=-\nabla_i\phi_{ji}$. Note that for $\alpha=2$ we recover the Newtonian gravity,  for $\alpha=-1$ Hooke's harmonic force, and finally $\alpha=1$ is the additive\footnote{Recently, \cite{Mil10} proposed a quasi-linear formulation of MOND called QuMOND. We stress that the $\alpha=1$ case studied here is not QuMOND.} analogue of the dMOND regime in which the MOND force behaves qualitatively as $1/r$. Some authors (e.g. \citealt{chava08}, \citealt{bou10}) introduce the nomenclature of {\itshape weak} long-range interactions and {\itshape strong} long-range interactions depending on whether the force vanishes or diverges for $r\to+\infty$;
here we label them as {\itshape gravity-like} ($\alpha>0$) or {\itshape harmonic oscillator-like} ($\alpha\leq0$). The forces in eq.(1) can be also divided in two families depending on the confining nature of their potential, separated by the case $\alpha=1$, when the potential diverges for zero and infinite separation. For $\alpha\geq1$ the total energy of a particle may be positive or negative, while it is always positive for $\alpha<1$ (having assumed zero potential energy for a system collapsed at the origin). Particles can escape only from systems with $\alpha>1$, while bound particles have negative energies.\\

In order to keep track of the process of virialization, for each simulation we follow the evolution of the virial ratio 
\begin{equation}
\eta=\frac{2K}{|W|},
\end{equation}
\begin{figure*}
\includegraphics[width=0.9\textwidth]{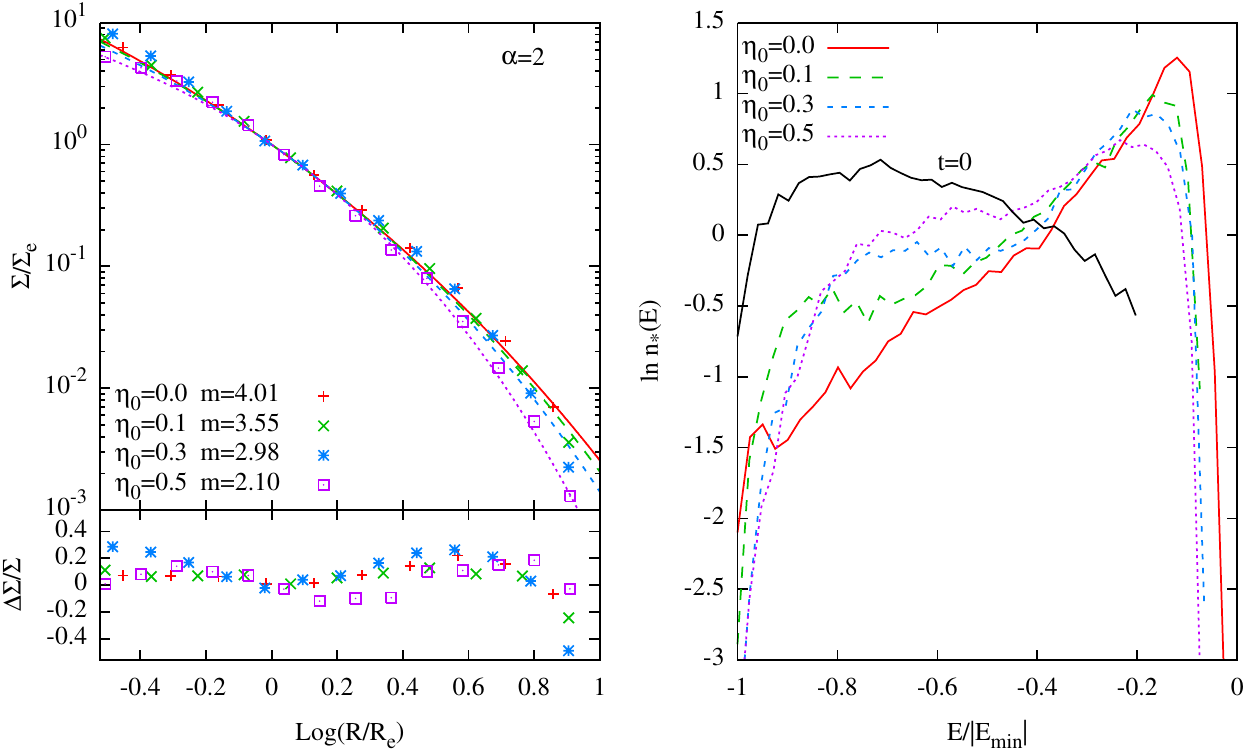}
\caption{Newtonian ($\alpha=2$) tests for Plummer initial conditions with $0\leq\eta_0\leq0.5$. Left: projected density profiles for the final states at 50 $t_*$ (dots), and their best-fit S\'ersic profiles (lines); residuals are also shown. Right: normalized differential energy distribution $n_*(E)=n(E)/N$ for the final states. The black solid line represents $n_*(E)$ for the initial condition with $\eta_0=0$; note that $E_{{\rm min}}$ depends on the specific realization.}
\label{test}
\end{figure*}
where 
\begin{eqnarray}\label{cinetica}
K=\sum_{i=1}^{N}\frac{m_i v_{i}^{2}}{2},
\end{eqnarray}
is the total kinetic energy of the system, and 
\begin{eqnarray}\label{viriale1}
W=-\int \rho(\mathbf{x})<\mathbf{x},\nabla\phi>d^3\mathbf{x}=\sum_{j\neq i=1}^N m_i<\mathbf{x}_i,\mathbf{a}_{ji}>
\end{eqnarray}
is the virial function, so that the virial theorem reads $2K=-W.$ Note that the condition $j\neq i$ in eq. (\ref{viriale1}) is required for $\alpha>0$, however this condition can be extended without loss of generality also to $\alpha\leq 0$, due to the vanishing of the self force. Note also that convergence of the integral in eq.(\ref{viriale1}) requires  $\alpha<3$. 
We recall that in general $W$ is {\it not} the total potential energy
\begin{eqnarray}
U=\frac{1}{2}\int\rho(\mathbf{x})\phi(\mathbf{x})d^3\mathbf{x}=\frac{1}{2}\sum_{j\neq i=1}^Nm_i\phi_{ji},
\end{eqnarray}
but for $\alpha\neq1$ it is proportional to it, being  
\begin{eqnarray}\label{viriale3}
W=(\alpha-1)U.
\end{eqnarray}
Note that $U=0$ when the system is dispersed at infinity for $\alpha>1$, while $U=0$ when the system is collapsed at the origin for $\alpha<1$. In the $\alpha=1$ case, the reference state must be fixed with the particles at finite (but non zero) separation.
The case $\alpha=1$, corresponding to the logarithmic potential (i.e. to dMOND-like force), is peculiar, as from eqs. (\ref{eq1}) and (\ref{viriale1}) one obtains\footnote{Note that in the case of a continuous density distribution with $\alpha=1$ and total mass $M$, $W=-GM^2/2$, and this is not the limit of eq.(8) for $N\rightarrow\infty$ and $m=M/N$.}
\begin{eqnarray}
W=-\frac{G}{2}\sum_{i\neq j=1}^N m_im_j,
\end{eqnarray}
i.e. $W$ remains constant during the virialization. Remarkably, it can be shown analytically that the virial function is time-independent also in dMOND, even though the field equation is not linear and the force is in general neither radial nor strictly proportional to $1/r$, (N07a; for the special case of spherical systems see \citealt{spergel}). As the total energy $E=K+U$ is conserved for all values of $\alpha$, oscillations in  $K$ are always associated with oscillations in $U$, so that for $\alpha\neq 1$ eq. (\ref{viriale3}) and the Lagrange-Jacobi identity $\ddot I=2(2K+W)$ (e.g. \citealt{ciotti2000}) show that the time dependence of the moment of inertia of the system is due to the combined effects of $K$ and $U$. For $\alpha=1$ instead only the kinetic energy changes during the virialization. Finally, for a dissipationless collapse, starting from cold initial conditions ($K_{{\rm in}}=0$), it is easy to prove that the value of the equilibrium wirial function ($W_{{\rm fin}}$) is related to the initial potential energy $U_{{\rm in}}$ by
\begin{eqnarray}\label{finale}
W_{{\rm fin}}=\frac{2W_{{\rm in}}}{3-\alpha}=\frac{2(\alpha-1)(U_{{\rm in}}-K_{{\rm fin}}^e)}{3-\alpha}, \quad (\alpha\neq1).
\end{eqnarray}
where $K_{{\rm fin}}^e$ is the asymptotic energy of the possible escapers. In the identity above it is assumed that the escapers are fully dispersed, i.e., that their gravitational energy is small, and that $W_{{\rm fin}}$ is the virial function of the remnant. Not also that in the cases with $\alpha<1$ no escape is possible, so that eq.(\ref{finale}) holds rigorously with $K_{{\rm fin}}^e=0$ (provided equilibrium could be attained), and $W_{{\rm fin}}$ refers to the whole system.
We used eq.(\ref{finale}) as a test for the simulations.
\begin{figure}
\begin{center}
\includegraphics[width=\columnwidth]{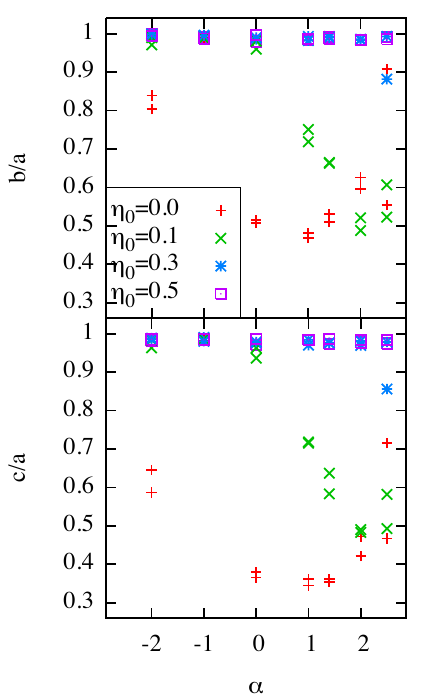}
\end{center}
\caption{Final axial ratios as function of the force exponent $\alpha$, for Hernquist and Plummer initial conditions with different values of $\eta_0$. Note how for each value of $\eta_0$ the results are almost identical independently of the initial density profile. In the harmonic oscillator case ($\alpha=-1$) the systems retain their spherical shape.}
\label{bm}
\end{figure}
\section[]{The simulations}
\subsection[]{The $N-$body code}
In order to compare the process of virialization of systems with different values of $\alpha$, we introduce the time-scale $t_*$ from the relation
\begin{eqnarray}\label{norm}
\frac{GMt_{*}^2}{2r_{*}^{\alpha+1}}=1,
\end{eqnarray}
where $r_{*}$ (the length scale) is the half-mass radius of the density distribution at $t=0$, and $M=Nm$ is the total mass of the system (see also DCC11); the natural velocity scale becomes  
\begin{eqnarray}\label{vnorma}
v_*\equiv\frac{r_*}{t_*}.
\end{eqnarray}
Setting $\mathbf{x}\equiv\mathbf{r}/r_*$ and $\tau\equiv t/t_*$, the dimensionless equations of motion for the particle $i$ become
\begin{eqnarray}\label{eqadim}
\frac{{\rm d}^2\mathbf{x}_i}{{\rm d}\tau^2}=-\frac{2}{N}\sum_{j\neq i=1}^N\frac{\mathbf{x}_i-\mathbf{x}_j}{||\mathbf{x}_i-\mathbf{x}_j||^{\alpha+1}}.
\end{eqnarray}
We note that, as all the presented results are scaled to the dynamical time $t_*$, the specific value of $G$ does not affect the conclusions. 
The code used for the simulations is a direct code, with the consequent limitations on the number of particles that can be used: as a rule we adopt $N=25000$. In the case of gravity-like forces ($\alpha>0$), the divergence of the force when the interparticle separation tends to zero is cured with the introduction of the softening length $\epsilon$ (e.g. \citealt{DEH01}), so that eqs.(\ref{eq1})-(\ref{eqp}) are replaced by their softened expressions
\begin{equation}\label{eqsofta}
{\bf a}_{ji}^{{\rm soft}}=-G m_j\times{\mathbf{r}_i-\mathbf{r}_j\over\left(||\mathbf{r}_i-\mathbf{r}_j||^2+\epsilon^2\right)^{\frac{\alpha +1}{2}}},
\end{equation}
\begin{equation}\label{eqsoftpot}
\phi_{ji}^{{\rm soft}} =G m_j\times \cases{
 \displaystyle{
 {\left(||\mathbf{r}_i-\mathbf{r}_j||^2+\epsilon^2\right)^{\frac{1-\alpha}{2}}\over1-\alpha}, \quad \alpha\neq1;}\cr
  \ln\sqrt{||\mathbf{r}_i-\mathbf{r}_j||^2+\epsilon^2}, \quad \alpha=1.
}
\end{equation}
The optimal $\epsilon$ is chosen as follows: the density profile of the initial conditions is divided in spherical shells of radius $R_i$ and thickness $\delta R_i$, and the minimum inter-particle distance $d_i$ within each shell is computed. A value $\epsilon_i$ is obtained by comparing the acceleration $a^{{\rm soft}}(d_i,\epsilon_i)$ of a pair with separation $d_i$ with ${a}^{{\rm soft}}_i(\epsilon_{i})$, the acceleration of a random particle of the shell due to all the other particles of the system. The value of $\epsilon_i$ is chosen so that ${a}^{{\rm soft}}_i(\epsilon_{i})=a^{{\rm soft}}(d_i,\epsilon_i)$. As optimal $\epsilon$ we take the maximum $\epsilon_i$. We verified that with such choice, independently of $\alpha$, the softened acceleration of a particle at large distance from the centre of mass of the system differs by less than 0.01\% from the non-softened acceleration. The equations of motion are integrated using a standard second order leapfrog method both with constant and adaptive timestep $\Delta t$. In the second case $\Delta t={\rm min}(\Delta t_i)$, where $\Delta t_i={\rm min}(\Delta t_{1i},\Delta t_{2i},\Delta t_{3i})$, and 
\begin{eqnarray}
\Delta t_{1i}\equiv\frac{||\Delta\mathbf{r}_i||}{||\Delta\mathbf{v}_i||},\quad\Delta t_{2i}\equiv\frac{m_i||\Delta\mathbf{r}_i||^2}{||\Delta\mathbf{J}_i||},\quad
\Delta t_{3i}\equiv\sqrt{\frac{m_i||\Delta\mathbf{r}_i||^2}{|\Delta E_i|}}.
\end{eqnarray}
$\Delta\mathbf{r}_i$, $\Delta\mathbf{v}_i$, $\Delta E_i$ and $\Delta\mathbf{J}_i$ are the variation of position, velocity, energy and angular momentum of the $i\rm{th}$ particle in the previous timesteps.
\subsection[]{Initial conditions}
In the present exploration, the force exponent spans the range $-5/2\leq\alpha\leq5/2$, and the initial conditions are characterized by values of the virial ratio $0\leq\eta_0=2K_{{\rm in}}/|W_{{\rm in}}|\leq0.5$. The $N$ particles are distributed in space with a standard rejection technique. In the first family, used to study the evolution of cuspy initial conditions, we adopt the \cite{her} density profile 
\begin{equation}
\label{her}
\rho(r)=\frac{Ma}{2\pi r(r+a)^3},
\end{equation}
where $M$ is the total mass, $a=r_*(\sqrt{2}-1)$ is the scale radius,
and $r_*$ is the half mass radius.
In the second family of initial conditions, characterized by a flat core, we use the \cite{Plu} density profile
\begin{equation}
\label{plum}
\rho(r)=\frac{3Ma^2}{4\pi\left(r^2+a^2\right)^{5/2}};
\end{equation}
in this case $a=r_*\sqrt{2^{2/3}-1}.$
We then extract the initial velocity of the particles from a position-independent Gaussian distribution, with the velocity dispersion tuned as to obtain the desired initial virial ratio $\eta_0$. The results are independent on the values of $M$ and $a$, which do not appear in the dimensionless equations of motion (equation 12), so for each of the two families we explore the two dimensional parameter space defined by the pair ($\alpha, \eta_0$). We note that similar initial conditions have been used recently for the numerical study of violent relaxation in Newtonian gravity (e.g. \citealt{visbal} and \citealt{fsl13}). All the simulations presented in this paper were performed on a cluster of LINUX HP\textregistered  Z700 workstations, and each run (on a single processor) lasts for $\approx4$ days when extended up to 50 $t_*$.
\begin{figure*}
\begin{center}
\includegraphics[width=0.95\textwidth]{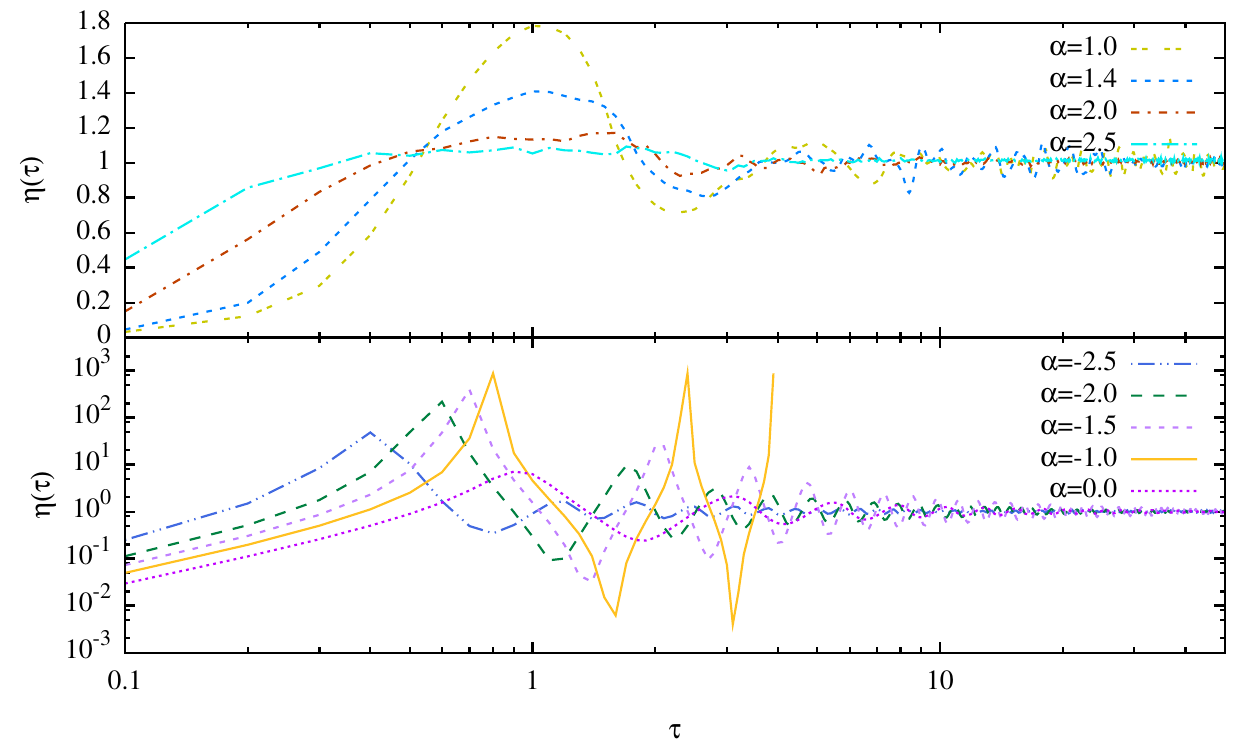}
\end{center}
\caption{Time evolution of the virial ratio $\eta$ for cold ($\eta_0=0$) Hernquist initial conditions with different values of $\alpha$. Upper panel: gravity-like forces ($\alpha>0$). Lower panel: harmonic oscillator-like forces ($\alpha\leq0$). For clarity, the unrelaxing case of the harmonic force ($\alpha=-1$, solid line) has been plotted only up to the third peak of $\eta$.}
\label{virialeg}
\end{figure*}
\subsection[]{Analysis of the numerical outputs}
The numerical outputs are used not only to study the dependence of the virialization process on the force law but also to investigate the structural and dynamical properties of the final states. We assume that the system has reached its final state when the amplitude of the oscillations of $\eta$ becomes smaller than $10^{-4}$ (which typically occurs at $t\approx30t_*$). Following \cite{NLC02} and \cite{MZ97}, we compute the second order tensor\footnote{Notice that $I_{ij}$ is {\it not} the inertia tensor, which is given instead by ${\rm Tr}(I_{ij})\delta_{ij}-I_{ij}$.}
\begin{equation}
I_{ij}\equiv m\sum_{k=1}^N r_i^{(k)}r_j^{(k)}
\end{equation}
for the particles inside the sphere of radius $r_{85}$ containing the 85\% of the total mass of the system, where $r_i$ are the Cartesian components of the position vector in the reference frame with origin in the centre of mass. 
The matrix $I_{ij}$ is diagonalized iteratively, requiring that the percentage difference of the largest eigenvalue between two iterations to be smaller than $10^{-3}$. This procedure requires on average 10 iterations, and we call $I_1\geq I_2\geq I_3$ the three eigenvalues. We finally apply a rotation to the system in order to have the three eigenvectors oriented along the coordinate axes. For of a heterogeneous ellipsoid of semiaxes $a,b$ and $c$, we would obtain $I_1=Aa^2$, $I_2=Ab^2$ and $I_3=Ac^2$, where $A$ is a constant depending on the density profile. Consistently, for the end-products we define $b/a=\sqrt{I_2/I_1}$ and $c/a=\sqrt{I_3/I_1}$, so that the ellipticities in the principal planes are $\epsilon_1=1-\sqrt{I_2/I_1}$ and $\epsilon_2=1-\sqrt{I_3/I_1}.$\\

It is known (see e.g. \citealt{vanAlbada82}, \citealt{LMS91}, \citealt{bert05}, N06a) that the end products of Newtonian dissipationless collapses starting from cold initial conditions (and also of MOND dissipationless collapses, see N07a) have surface density profiles  
well described by the S\'ersic law
\begin{equation}\label{sersic}
\Sigma(R)=\Sigma_e e^{-b\left[\left(\frac{R}{R_e}\right)^{1/m}-1\right]},
\label{serse}
\end{equation}
where  $b\simeq2m-1/3+4/405m$ (\citealt{ciotti99}), and $\Sigma_e$ is the projected mass density at effective radius $R_e$, the radius of the circle containing half of the projected mass. 
In practice, in our analysis we circularize the projected density in the 3 principal planes of the virialized systems, and by particle count we determine the corresponding pair $(R_e,\Sigma_e)$, so that from eq. (\ref{serse}) we obtain the best-fit $m$. In this way for each simulation we determine 3 sets of ($\Sigma_e$, $R_e$, $m$) and we then chose randomly one of them, being the others in general qualitatively similar.\\

In the same spirit as previous works (DC09, DCC11) we focus on different indicators of relaxation, such as the evolution of the virial ratio and of the phase-space sections ($r,v_r$). We also study the final differential energy distribution $n(E)$, defined by the relation
\begin{equation}\label{ne}
\int_{E_{{\rm min}}}^{E_{{\rm max}}} n(E)dE=N
\end{equation}
(e.g. \citealt{bt2008}). Finally, we construct the so-called pseudo-phase-space density of the final states and we check if they obey the GDSAI, as described in Sect. 4.3.
\subsection[]{Testing the code}
As a first set of numerical experiments, we determined the optimum choice of the softening length and of the time step to be used in the simulations. Following the procedure described in Section 3.1, we found that, independently of $\alpha$, $\epsilon\simeq 10^{-2}r_*$ guarantees not only numerical accuracy of the results (with energy conservation better than 3\% at virialization in the worst cases, and usually better than the 0.5\%), but also acceptable computational times. In addition, comparing the evolution of collapses starting from identical initial conditions with adaptive or fixed timesteps, we found that a fixed timestep $\Delta t\simeq10^{-2}t_{*}$ guarantees a good balance between computational time and energy and total angular momentum conservation independently of the value of $\alpha$ and of the initial profile, so we adopt this criterion for all the simulations.\\

We tested our direct code in the Newtonian case against several well established results of numerical simulations of cold collapses obtained with the tree-code {\sc FVFPS}  (\citealt{fvfps}), as well as in the Newtonian limit with the particle-mesh MOND code {\sc N-MODY} (\citealt{NMODY}). In particular, we performed collapses for different initial density profiles and values of the virial ratio. We fit the final surface density profile with the S\'ersic law over the radial interval $0.3R_e-10R_e$. As shown in Fig. \ref{test} (left panel) the resulting values of $m$ range from $\simeq2$ for the hottest initial condition ($\eta_0=0.5$) to $\simeq4$ for the cold collapse ($\eta_0=0$). Over the radial range here considered, the final profiles are indistinguishable from those obtained by N06a for comparable values of the initial virial ratio ($0\leq\eta_0\leq0.2$). As can be seen from Fig. \ref{bm} the final states, for both Hernquist and Plummer initial profiles, are roughly spherical for $\eta_0>0.1$ and prolate ($c/a \simeq b/a \simeq 0.5$) for $\eta_0\leq0.1$, consistent with the results of N06a, which indicates that colder systems are more prone to undergo instabilities that perturb significantly their initial shape. Finally, the tests confirm that after virialization $n(E)$ is well described over a broad range of energies by an exponential function. Remarkably, slightly bimodal final differential energy distributions characterize the systems starting from hotter initial conditions (Fig. \ref{test}, right panel) as seen in analogous plots of previous papers (figure 2 in N06b and figure 8 in \citealt{LMS91}). 
\begin{figure*}
\begin{center}
\includegraphics[width=0.8\textwidth]{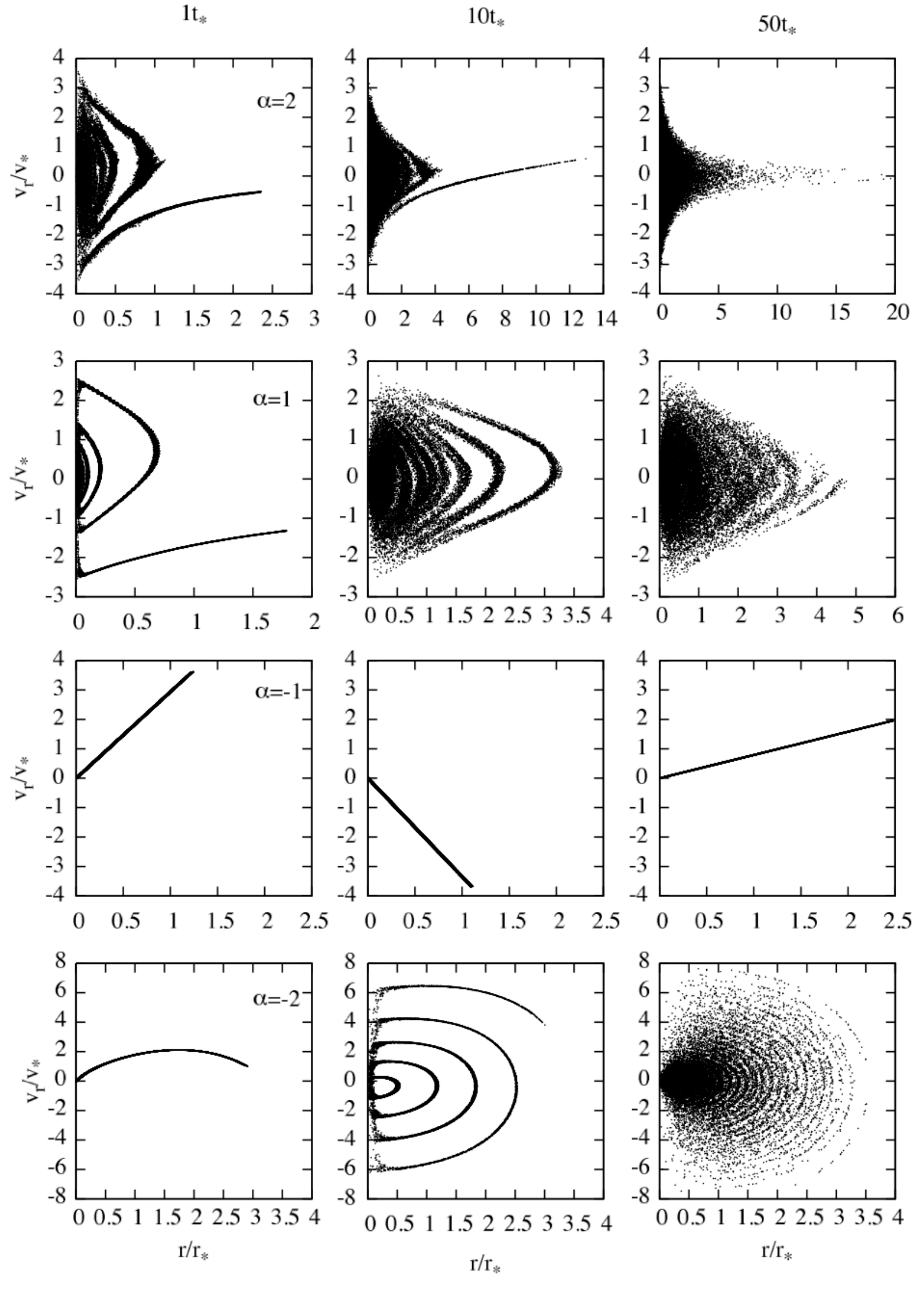}
\end{center}
\caption{The evolution of phase-space sections $(r, v_r)$ for Hernquist initial conditions with $\eta_0=0$, and $-2\leq\alpha\leq2$. As expected, no mixing is acting on the harmonic-oscillator case ($\alpha=-1$, where the straight line rotates clockwise and each point of it describes similar ellipses), while mixing appears again in the superharmonic case ($\alpha=-2$). The characteristic time and length scales, $t_*$ and $r_*$, are related as given by eq. (10).}
\label{fasi}
\end{figure*}
\section[]{Results}
\subsection{The relaxation process}
One of the motivations of this study is to elucidate the reason of the long relaxation time (in units of their dynamical time) of dMOND systems when compared to the same quantity for Newtonian systems. A simple measure of the relaxation effectiveness can be obtained considering the number and the decay rate of the major oscillations of the virial ratio $\eta$. In Fig. \ref{virialeg} we show the
evolution of $\eta$ as a function of the dimensionless time $\tau$ for cold ($\eta_0=0$) Hernquist initial conditions  with different values of $\alpha$. The top panel illustrates the evolution for the family of gravity-like forces. The similarity with the results obtained with the shell models (Fig.1 in DCC11) is remarkable: a decrease of $\alpha$ leads to a higher value of $\eta$ at the first peak, and to a longer series of virial oscillations of decreasing amplitude. Curiously, for $\alpha=1$ the amplitude of the virial oscillations increases again at large times, similarly to dMOND collapses (N07a, \citealt{CNL07}), confirming that dMOND behaves qualitatively as the $1/r$ force when considering a system not deviating too much from spherical shape, reinforcing the previous result of the long relaxation times of $N-$body systems governed by the non-linear field equation of MOND. We interpret the large peak values of $\eta$ for small $\alpha$ as due to the fact that the force is stronger on large scales, and that the systems with low $\alpha$ collapse more as a whole, consistently with the force being more similar to the harmonic oscillator case. It is important to recall that in the Newtonian case a spherical homogeneous shell does not exert any force inside, while inside a shell the force is directed outwards for $\alpha>2$, and the opposite happens for $\alpha<2$ (e.g. DCC11). Therefore, when $\alpha>2$ the external regions act {\it against} the collapse, while for $\alpha<2$ also the external regions of the system contribute more and more to the collapse.\\   

The results for collapses driven by harmonic-like forces are shown in the bottom panel of Fig. \ref{virialeg}. Note how the virial ratio oscillates with peak values of $\eta$ significantly larger, and more regular oscillations than in the gravity-like cases. As expected, in the $\alpha=-1$ case no relaxation takes place, since the whole system behaves as a single harmonic oscillator (e.g. \citealt{LB82}, see also eq. (\ref{eqadim})). In particular, in a system of harmonic oscillators starting at rest all the particles cross the centre simultaneously, so that $|W|\rightarrow0$ while $K\rightarrow-E$, causing $\eta$ to diverge. For the reasons described above, in the super-harmonic case ($\alpha<-1$), the first peak of $\eta$ is reached at earlier times for decreasing $\alpha$. However, the peak values decrease, due to phase mixing which forbids all the particles to cross the centre simultaneously. Cases with $\eta_0$ slightly larger than $0$, and for the Plummer initial conditions, are not shown, being qualitatively the same. In general, large values of $\eta_0$ 
\begin{figure*}
\begin{center}
\includegraphics[width=0.9\textwidth]{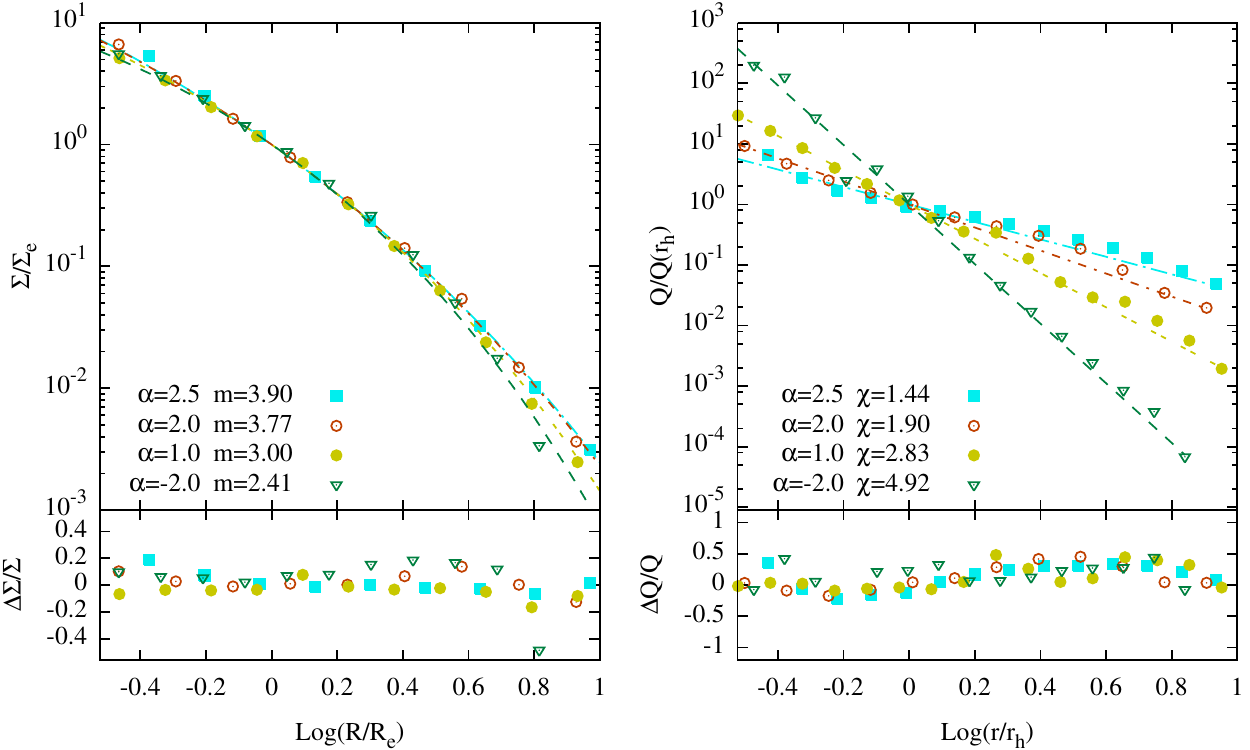}
\end{center}
\caption{Left: projected density profiles of the end-products of Hernquist initial conditions with $\eta_0=0$ and different values of $\alpha$ (dots). Dashed lines are their S\'ersic best fit. Right: the  associated pseudo phase-space density. Residuals with respect to the best fit are also shown.}
\label{beta0}
\end{figure*}
correspond to small amplitudes of the first peak of $\eta$.\\

The long lasting virial oscillations for forces close to the harmonic oscillator are associated with a poorer mixing in phase space. Such behavior is evident from the evolution in the phase-space section ($r,v_r$) defined as radial position and radial velocity. In Fig. \ref{fasi} we show snapshots of the phase-space at  1, 10 and 50 $t_*$ for $\alpha=$-2, -1, 1, 2. Consistently with the findings of DCC11 (where the narrower interval of $1\leq\alpha\leq2$ was studied), larger values of $\alpha$ show a more efficient phase mixing with respect to systems with $\alpha\leq0$. Again, the similarity with the plots in N07a (their figure 4) and Ciotti et al. (2007, their figures 2 and 3) is remarkable. One may speculate that the coherent structures in phase space that persist at large times (in units of $t_*$) are akin to the so-called {\it phase-space holes} reported by some authors (\citealt{mineau90},  \citealt{joyce10} and \citealt{TELE11}) in the context of the one dimensional infinite sheet model, where the mixing is quite poor. It must be pointed out that the poorer mixing in Newtonian gravity in lower dimensions is essentially due to the smaller number of degrees of freedom that are involved in the relaxation rather than a different exponent in the force law (see e.g. \citealt{kan89}). As already remarked, the efficiency of phase mixing is non monotonic with $\alpha$, with no mixing for $\alpha=-1$.
\subsection{Structural properties of the end products}
The triaxiality of the final states of the collapses is shown in Fig. \ref{bm} where we plot the values of the axial ratios $b/a$ and $c/a$ of the end products at 50 $t_{*}$, for representative values of $\alpha$ and for increasing values of $\eta_0$. In general, the Newtonian behavior is confirmed, in the sense that at fixed $\alpha$ triaxiality is more pronounced for small values of $\eta_0$, for both Plummer and Hernquist initial conditions. Again, the only exception is the $\alpha=-1$ force, when the systems retain their spherical shapes, consistently with their orbital structure. For given $\eta_0$, the triaxiality as a function of $\alpha$ shows characteristic non-monotonic trend especially visible in the bottom panel of Fig. \ref{bm} for $\eta_0=0$ and $\eta_0=0.1$. For decreasing $\alpha$ the flattening increases, reaches a maximum, and then decreases again. The maximum values of the triaxiality (the minimum values of $b/a$ and $c/a$) are obtained for $1\leq\alpha\leq2$ when $0\leq\eta_0\leq0.1$, with a quite clear correlation between $\alpha$ and $\eta_0$. Remarkably, no models are found to be flatter than an elliptical galaxy E7 (i.e. $c/a\geq0.3$), thus leading to conjecture that this limit may hold more generally than just in Newtonian gravity.\\

In analogy with the case of Newtonian collapses, we fitted the final projected density profile with the S\'ersic law as described in Sect. 3.3. In general, we find that the S\'ersic law provides a good description of most of the final states for both Plummer and Hernquist initial conditions. More quantitatively, in the left panel of Fig. \ref{beta0} we show the end state of the collapse of a perfectly cold Hernquist initial condition. The main result is a quite well defined dependence of the S\'ersic index $m$ on $\alpha$, 
\begin{figure*}
\begin{center}
\includegraphics[width=\textwidth]{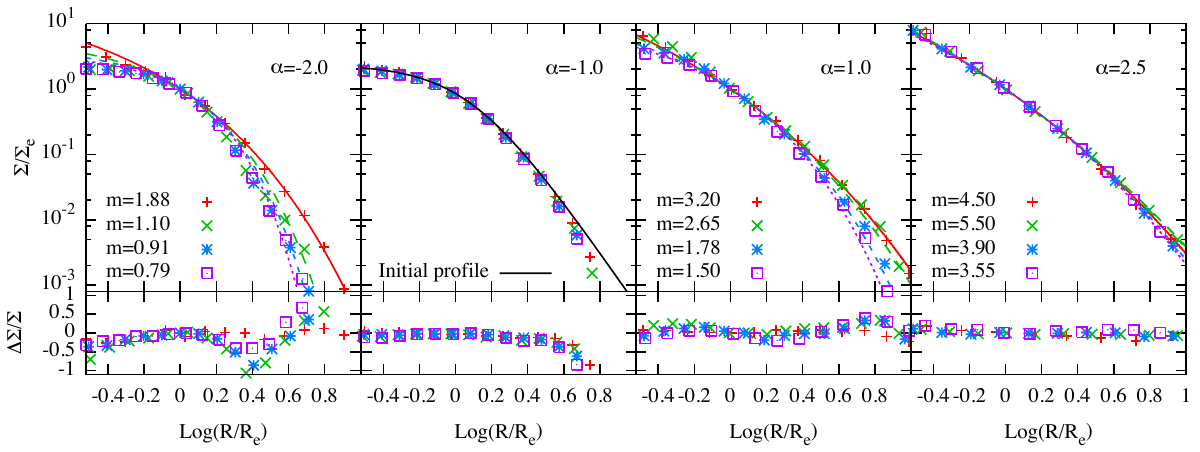}
\end{center}
\caption{Projected density profiles of the end products at 50 $t_*$ (points) and their S\'ersic best fits (lines) of Plummer initial conditions for $\alpha=-2,-1,1,2.5$ and different values of $\eta_0$. In the $\alpha=-1$ case (harmonic force), the projected density profile is scale invariant, as expected. The symbols are the same as in Fig. 1.}
\label{nesersic2}
\end{figure*}
with large values of $m$ associated with large values of $\alpha$. In practice, gravity-like forces produce more peaked density profiles than harmonic like forces. The $\alpha=-1$ case is not shown, as the profile collapses and expands self-similarly. For other values of $\alpha$, percentual deviations of the data from the fits are well within the 20\%. The model with the largest deviations is the superharmonic one with $\alpha=-2$ and $m\simeq2.4$. The dependence of the final states on $\eta_0$ is shown in Fig. \ref{nesersic2}. In general, hotter initial conditions lead to smaller values of $m$, independently of $\alpha$ (for the Newtonian case see Fig. \ref{test}, left panel). Moreover, the largest deviations from the best fit are again produced by the superharmonic $\alpha=-2$ force, while the $\alpha=1$ case is remarkably similar to the dMOND results of N07a.\\

\begin{figure}
\begin{center}
\includegraphics[width=\columnwidth]{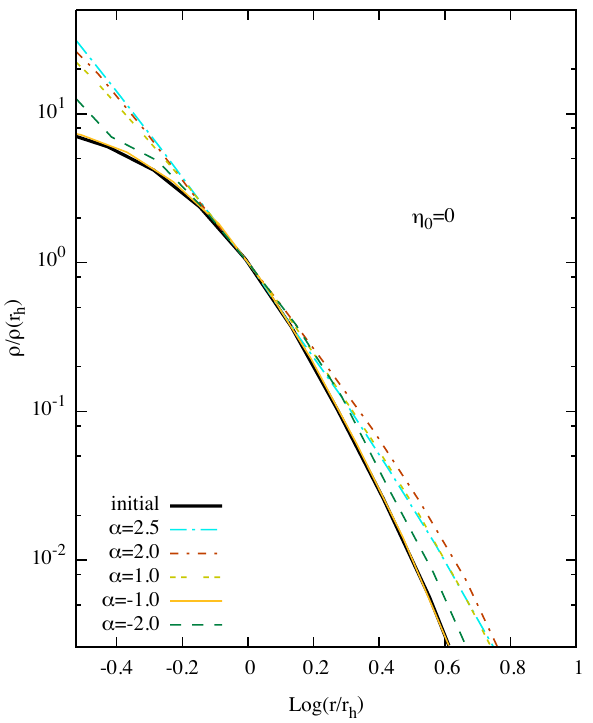}
\end{center}
\caption{Normalized angle averaged three dimensional density profiles of the 
end products of Plummer initial conditions (black solid line) with 
$\eta_0=0$, and different values of $\alpha$. Radii are normalized to the 
volumetric half-mass radius of the final state.}
\label{figure7}
\end{figure}

For completeness, in Fig. \ref{figure7} we also show the three dimensional 
(angle-averaged) density profiles of the end-products of cold ($\eta_0=0$) 
Plummer initial conditions. As apparent, and in agreement with the 
projected density profiles, higher valiues of alpha corresponds to more 
peaked final density profiles, while for $\alpha=-1$ the normalized profile 
does not change.
\subsection{The differential energy distribution, the pseudo phase-space density and the GDSAI of the end products}
In addition to their structural properties, the virialized final states of collapses are usually also studied from the point of view of the phase-space properties. Here, following a well established approach, we focus on their differential energy distribution, on the radial trend of the so-called pseudo phase-space density, and finally on the density-slope inequality.\\ 

\begin{figure*}
\begin{center}
\includegraphics[width=\textwidth]{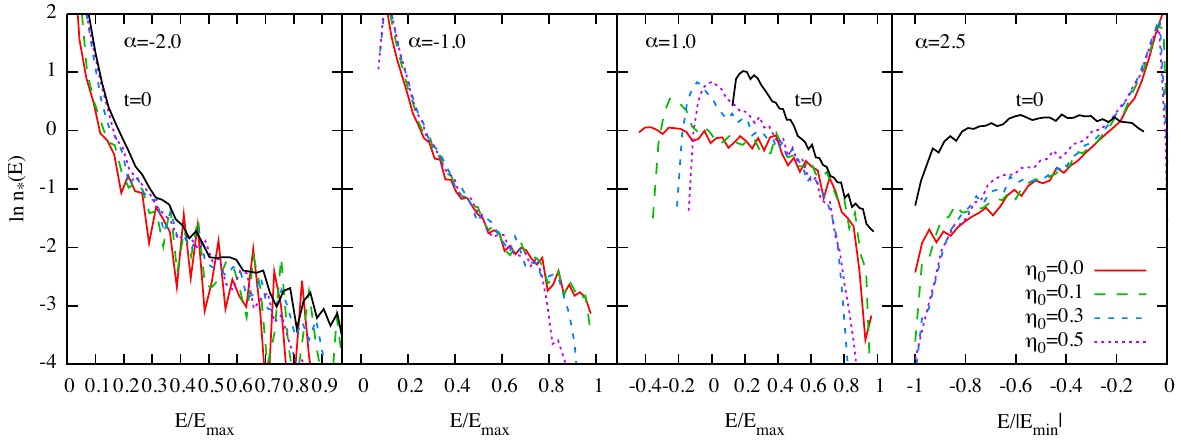}
\end{center}
\caption{Normalized differential energy distribution $n_*(E)=n(E)/N$ of the end products of Plummer initial conditions for different values of $\eta_0$ and $\alpha=-2,-1,1,2.5$. The reference energy is $E_{{\rm max}}$ or $|E_{{\rm min}}|$ if $\alpha\leq 1$ or $\alpha>1$ respectively. The Newtonian case ($\alpha=2$) is shown in Fig.1. The heavy solid lines represent the $n_*(E)$ for the initial condition with $\eta_0=0$. Hernquist initial conditions lead to very similar energy distributions.}
\label{ne1and25}
\end{figure*}

In Fig. \ref{ne1and25} we show the final differential energy distribution $n(E)$ for $-2\leq\alpha\leq2.5$, and for different values of $\eta_0$. Each distribution is normalized to the total number of particles, 
and the energy range to $E_{{\rm max}}$ (for $\alpha\leq1$) and to $|E_{{\rm min}}|$ (for $\alpha>1$) of the final states. The initial conditions with $\eta_0=0$ are represented by the heavy solid lines. The first important and general feature is that $n(E)$ is peaked at high energies for $\alpha>1$ (see also Fig. 1, right panel, for the Newtonian case). In practice, the virialized final states of systems with gravity-like 
forces allowing for escape are mainly supported by loosely bound particles (e.g., \citealt{bt2008}, \citealt{binney82}, \citealt{ciotti91}) corresponding to particles in the outer regions. Note also how $n(E)$ evolves significantly due to relaxation, with major changes at high energies. For harmonic-like forces the situation is different, and very little evolution is found. Both the initial conditions and the final states have a $n(E)$ distribution peaked at low energies. Of course, consistently with the extraordinary nature of the harmonic force, the $n(E)$ for the $\alpha=-1$ case is not evolving (barring numerical fluctuations). In general, different values of $\eta_0$ in the range explored do not affect significantly the shape of the $n(E)$ with the exception of $\alpha=1$ forces. The systems with $\alpha=1$ show an intermediate behavior, with a trend similar to dMOND collapses, (See N07a, Fig. 5 therein). For this latter case it is apparent how decreasing values of $\eta_0$ tend to populate the external regions of the final systems. As discussed in the Introduction, a specific feature of the $n(E)$ obtained in Newtonian collapses is the exponential shape over some energy range. Here, due to the energy sign associated with the value of $\alpha$ (see discussion in Sect. 2) we consider the function
\begin{equation}\label{expdist} 
n(E)=A\times\cases{
 \displaystyle
  e^{-\theta|E|}, \quad \alpha>1;\cr
  e^{-\theta E}, \quad \alpha\leq1;
}
\end{equation}
where $\theta$ and $A$ are an inverse (positive) temperature and a normalization factor respectively. It is apparent that for $\alpha<1$ the shape of $n(E)$ cannot be described by a single-temperature exponential distribution, independently of the hotness of the initial conditions. Instead, for gravity-like forces with $\alpha\geq1$ a larger energy range exists over which $n(E)$ can be qualitatively described with an exponential function as in eq. (\ref{expdist}). The main difference in the gravity-like forces is between $\alpha=1$ case and the other case with $\alpha>1$ (see also Fig. 1): while in the forces allowing for escape $(\alpha>1)$ the exponential region is peaked towards high energies, in the $\alpha=1$ case the peak is at low energies, corresponding to the central regions. Interestingly, for $\alpha\geq1$ the trend between the S\'ersic index $m$ and the inverse temperature $\theta$ of the best fit $n(E)$, is qualitatively similar to what found by \cite{ciotti91} in the analysis of the Newtonian S\'ersic models. We finally note how the current $N-$body simulations produced final $n(E)$ much better described by an exponential distribution than in the shell model (DCC11), a natural consequence of a better energy exchange among the components of the system. For the final states of the systems with $\alpha>1$ we also considered the fraction of escapers (i.e. particles having positive energy, see also \citealt{joyce09} and \citealt{fsl13}). As expected, at fixed $\alpha$ and for given initial density profile, with low values of $\eta_0$ (i.e., cold initial conditions) there is a larger number of escapers. Also, for fixed density profile and $\eta_0$, the fraction of escapers is found to be weakly dependent on the value of $\alpha$, with values $\simeq3$\% for most cases and with the maximum value of $\simeq8$\% for perfectly cold Plummer model with $\alpha=2.5$.\\

Another property of interest, recently focus of several investigations, is the so-called pseudo phase space density (e.g. see \citealt{tana01}, \citealt{ab05}, \citealt{han}, \citealt{Lud10}, \citealt{bar12}, \citealt{sparre12}) defined as 
\begin{equation}
Q=\frac{\rho(r)}{\sigma^3(r)},
\label{defrhosigma}
\end{equation}
where $\rho(r)$ and $\sigma(r)$ are the angle-averaged density and velocity dispersion at radius $r$.
For Newtonian collapses, the numerical simulations have unequivocally shown that $Q$ is described quite well by a power-law
\begin{equation}\label{qpowrlaw}
Q\propto r^{-\chi},\quad\chi\simeq1.87.
\label{rhosigma}
\end{equation}
neither the origin of this relationship nor the dependence of this property on the initial conditions are yet fully understood despite the numerous efforts. On one side, the results of numerical simulations seem to point out to a remarkable robustness of eq.(\ref{qpowrlaw}): even though the power-law trend of $Q$ was initially considered a peculiarity of the NFW profiles (\citealt{NFW}), it is known that other profiles share this property (e.g. the family of self-consistent $f_{\infty}$ models, see \citealt{bert84}, \citealt{zocchi}). However, there are self-consistent equilibrium systems where $Q$ is not
a power-law (e.g. the Plummer sphere). Therefore it is natural to ask wether the power-law is a specific feature of violent relaxation in Newtonian gravity or its origin should be searched more in the physics of dissipationless collapse, independently of the force law involved. Here we are in the ideal position to address this question, and in fact the obtained results are quite significant. As can be seen in Fig. \ref{beta0} (right panel) 
a power-law trend for $Q$ is reproduced surprisingly well also in the case of non-Newtonian forces. For all the considered case (with the exception of  $\alpha = -1$), and independently of the initial density profile, at fixed $\eta_0$ the exponent $\chi$ increases for decreasing $\alpha$ and the function $Q$ steepens. We also found that at fixed $\alpha$, $Q$ steepens for decreasing $\eta_0$. These findings lead to conclude that a power-law radial dependence of $Q$ is more a consequence of violent relaxation than of the Newton gravity law.\\ 

Finally, we check if the (angle-averaged) Global Density Slope Anisotropy Inequality (GDSAI) is obeyed by the end products for different values of $\alpha$. Ciotti \& Morganti (2010ab), prompted by the important asymptotic result of \cite{anevans}, proved that a very large class of Newtonian stellar systems with positive phase-space distribution function, necessarily obey the inequality  
\begin{equation}
\gamma(r)\geq2\beta(r), \quad \forall r
\label{gammaxi}
\end{equation}
where 
\begin{equation}
\gamma=-\frac{{\rm d}\ln\rho}{{\rm d}\ln r}
\end{equation}
is the logarithmic density slope and 
\begin{figure*}
\begin{center}
\includegraphics[width=0.9\textwidth]{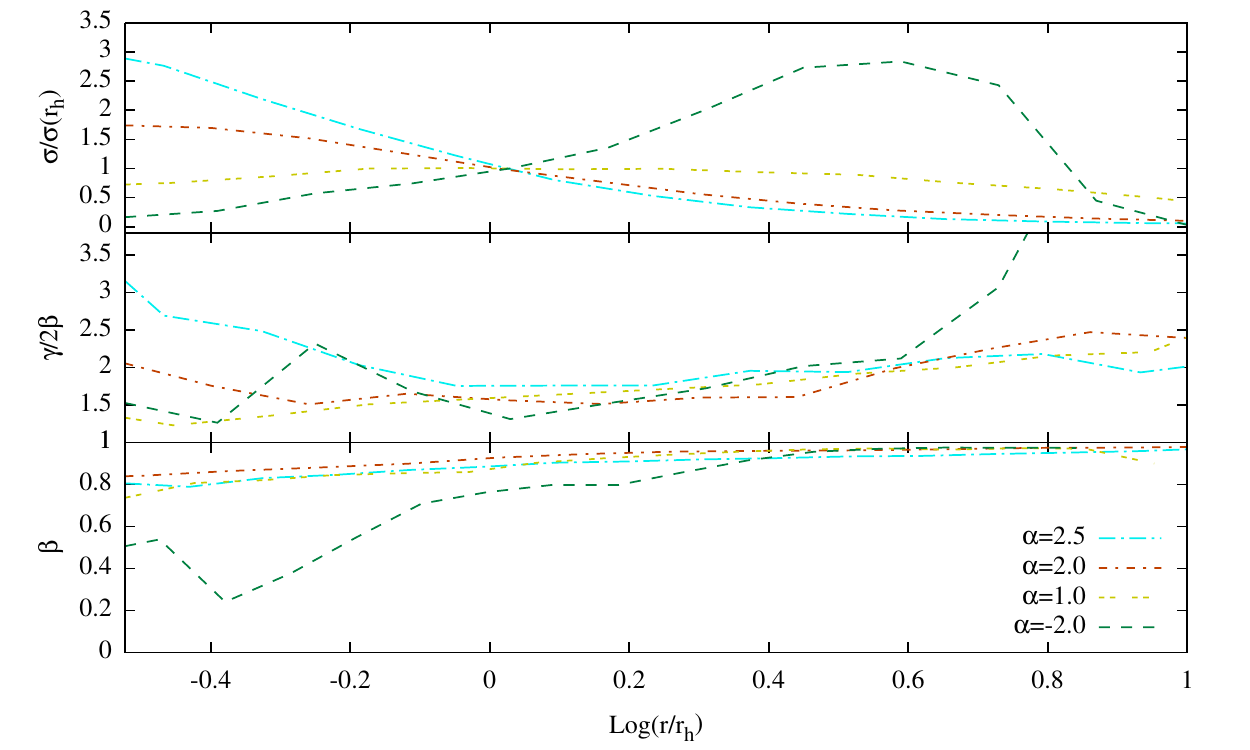}
\end{center}
\caption{Radial profiles of the angle-averaged velocity dispersion (top), GDSAI indicator (middle), and anisotropy parameter (bottom), for the final virialized states of cold ($\eta_0=0$) Hernquist initial conditions. $r_h$ is the volumetric half-mass radius of the final states. Similar trends are found also for larger values of $\eta_0$ and for Plummer initial conditions. Note the peculiar off-center maximum of $\sigma$ in the superharmonic force case.}
\label{gammas}
\end{figure*}
\begin{equation}
\beta=1-\frac{\sigma_t^2}{2\sigma_r^2}
\end{equation}
is the usual anisotropy parameter (\citealt{bt2008}). In the formula above $\sigma_r$ and $\sigma_t$ are the radial and tangential component of the velocity dispersion tensor respectively. In particular, \cite{CM10b} speculated about a possible universality the GDSAI, even though their analytical methods where unable to treat the cases with $\beta(0)>1/2$. Significant progress and clarification has been made in the subject (\citealt{vanhese11}, \citealt{an12}) and now the case of systems with {\it separable augmented density} is well understood: the GDSAI is obeyed by all separable systems with $\beta(0)\leq1/2$, while counterexamples exist for systems with $\beta(0)>1/2$. Much less is known about systems with non-separable augmented density, but numerical simulations in Newtonian gravity seem to suggest that also in general systems the GDSAI is usually satisfied. Here we analyzed the results of the simulations for different values of $\alpha$ and $\eta_0$. The trends of $\gamma$ and $\beta$, obtained using spherical averages and excluding the innermost regions (where discreteness effects dominate), revealed that the final states obey the GDSAI (with the obvious exception of the harmonic oscillator force). This is shown in Fig. \ref{gammas} where, even in presence of numerical noise, it is apparent that overall $\gamma\geq2\beta$, reinforcing the idea that the physical process leading to the establishment of the GDSAI may be independent of the specific force law considered. Finally, from the bottom panel of Fig. \ref{gammas}, it is also interesting to note how the final systems are significantly radially anisotropic in their outer regions, and become more and more isotropic near the centre, as also commonly found in Newtonian simulations.
 The $\alpha=-2$ case stands out as the less anisotropic and it is curious to recall that these systems are also those for which the S\'ersic law provides the less satisfactory description (Fig. \ref{nesersic2}).

\section{Discussion and conclusions}
As discussed in the Introduction, several theoretical arguments point toward the importance of elucidating the process of dissipationless collapse and virialization of $N-$body systems with additive interparticle forces proportional to $r^{-\alpha}$, a generalization of the Newtonian force. For this task we built a direct $N-$body code: preliminary results obtained with a shell model in spherical symmetry (\citealt{dicintio2009}, \citealt{dicintio2011}) appear to be confirmed by the present simulations. The main results can be summarized as follows.\\

 The relaxation process, independently of the initial density profile (Hernquist or Plummer), is characterized by a first phase of strong oscillations of the virial ratio, followed by a gentler phase of relaxation. For decreasing $\alpha$, the peak value of the virial ratio increases reaching a value formally infinite in the case of a perfectly cold collapse with $\alpha=-1$ (i.e. a system of harmonic oscillators), and then decreases again. This non-monotonic behavior is a consequence of the different degrees of phase mixing as a function of $\alpha$. Qualitatively, this effect can be understood by considering the force field inside a shell of matter for different values of $\alpha$. As expected, systems with $\alpha=-1$ do not relax due to their extraordinary orbital structure in which each particle behaves as an isolated harmonic oscillator (e.g. \citealt{LB82}).\\

With the obvious exception of the $\alpha=-1$ force, when the systems retain their initial spherical shape,  the final states are triaxial. As a rule triaxiality increases for colder initial conditions, similarly to what happens in Newtonian collapses. However, for fixed initial virial ratio $\eta_0$, triaxiality is not a monotonic function of $\alpha$: for decreasing $\alpha$ the flattening increases, reaches a maximum and then decreases again. The value of $\alpha$ for which the triaxiality is maximum depends on $\eta_0$, but it is almost independent on the initial density profile. Remarkably, no models are found to be flatter than an elliptical galaxy of type E7, independently of $\alpha$ and $\eta_0$.\\

In general the S\'ersic law provides a good description of most of the final states, with large S\'ersic index $m$ associated with large values of $\alpha$ (i.e. for gravity-like forces), and with cold initial conditions. Hotter initial conditions and harmonic-like forces produce density profiles characterized by smaller $m$. Moreover the quality of the S\'ersic fit deteriorates for low values of $\alpha$.\\

The differential energy distribution $n(E)$ of the final states shows two distinct behaviors, separated by the case $\alpha=1$. In particular, $n(E)$ is well described over a large range of energies by an exponential function peaked at high energies for gravity-like forces with $\alpha>1$. For $\alpha=1$, the final $n(E)$ is also exponential, but now the distribution is peaked at low energies (as in MOND simulations, N07a). Finally, when $\alpha<1$, very little evolution is found, and $n(E)$ remains peaked at low energies. Remarkably, for $\alpha\geq1$ the trend of the inverse temperature $\theta$ with the S\'ersic index $m$ is similar to what found for the Newtonian S\'ersic models, with $\theta$ increasing for decreasing $m$ (\citealt{ciotti91}).\\ 

We found that the pseudo phase-space density $Q=\rho/\sigma^3$ of the (angle-averaged) final states is described very well by a power law of radius $r^{-\chi}$, over a large radial range. In general, $Q$ steepens for decreasing $\alpha$, while at fixed $\alpha$ it flattens for increasing values of $\eta_0$. In addition we also found that the GDSAI holds (well within numerical uncertainties) for all the virialized end-states, and the amount of radial anisotropy tends to be higher for $\alpha>0$ (i.e. for gravity-like force) than for $\alpha<0$.\\

Overall the main conclusion of the present study are that several structural and dynamical features of the virialized states of cold and dissipationless collapses are not restricted to the special nature of the Newton law, but they appear to be more a property of the long range forces.  Among the radial $r^{-\alpha}$ forces, however, the gravity-like forces ($\alpha>0$) are those with the results more similar to the $1/r^2$ force. In addition, we found that the systems with interparticle force proportional to $1/r$ behave in many respects
as dMOND systems. In particular, we confirmed that 
the relaxation time (in units of the internal dynamical time) is longer for
$\alpha=1$ than for $\alpha=2$, and this is due to the force becoming 
more similar to the harmonic oscillator case ($\alpha=-1$) when the system oscillates forever. Future explorations, exploiting the similarity between $1/r$ and dMOND forces, will be focused on the study of radial orbit instability for $1/r^{\alpha}$ forces, in the line of the MOND study of \cite{NCL11}. 
\section*{Acknowledgements}
We thank Steen Hansen, Alberto Parmeggiani and the anonymous Referee for useful comments. LC and CN acknowledge financial support from PRIN MIUR 2010-2011, project ``The Chemical and Dynamical Evolution of the Milky Way and Local Group Galaxies'', prot.
2010LY5N2T. This material is based upon work supported in part by the National Science Foundation under grant No. 1066293 and the hospitality of the Aspen Center for Physics. 

\label{lastpage}

\begin{thebibliography}{99}

\bibitem[\protect\citeauthoryear{An \& Evans} {2006}]{anevans} An J.H. \& Evans N.W., 2006, ApJ, 701, 1500

\bibitem[\protect\citeauthoryear{An {\rm et al.}} {2012}]{an12} An J.H., van Hese E., Baes, M., 2012, MNRAS, 422, 652

\bibitem[\protect\citeauthoryear{Andredakis et al.} {1995}]{andre95} Andredakis Y. C., Peletier R. F., Balcells M.,  1995, MNRAS, 275, 874

\bibitem[\protect\citeauthoryear{Ascasibar \& Binney} {2005}]{ab05} Ascasibar Y.\& Binney J., 2005, MNRAS, 356, 872

\bibitem[\protect\citeauthoryear{Barber {\rm et al.}} {2012}]{bar12} Barber J. A., Zhao H., Wu X., Hansen S. H.,  2012, MNRAS, 424, 1737

\bibitem[\protect\citeauthoryear{Bertin {\rm et al.}} {2002}]{bert02} Bertin G., Ciotti L., Del Principe M.,  2002, A\&A, 386, 149

\bibitem[\protect\citeauthoryear{Bertin \& Stiavelli} {1984}]{bert84} Bertin G. \& Stiavelli M., 1984, A\&A, 137, 26

\bibitem[\protect\citeauthoryear{Bertin \& Trenti} {2003}]{bert03} Bertin G. \& Trenti M., 2003, ApJ, 584, 729

\bibitem[\protect\citeauthoryear{Bekenstein \& Milgrom} {1984}]{BM84} Bekenstein J. \& Milgrom M.,  1984, ApJ, 286, 7

\bibitem[\protect\citeauthoryear{Binney} {1982}]{binney82} Binney J., 1982, MNRAS, 200, 951

\bibitem[\protect\citeauthoryear{Binney \& Tremaine} {2008}]{bt2008} Binney J. \& Tremaine S., 2008 Galactic Dynamics, 2nd Ed. (Princeton University Press)

\bibitem[\protect\citeauthoryear{Bouchet {\rm et al.}} {2010}]{bou10} Bouchet F., Gupta S.. Mukamel D., 2010, Physica A, 389, 4389

\bibitem[\protect\citeauthoryear{Brandao \& de Araujo} {2012}]{BA12} Brandao C.S.S. \& de Araujo J.C.N., 2012, ApJ, 750, 29

\bibitem[\protect\citeauthoryear{Caon et al.} {1993}]{caon93} Caon N., Capaccioli M., D'Onofrio M., 1993, MNRAS, 265, 1013

\bibitem[\protect\citeauthoryear{Chavanis} {2008}]{chava08} Chavanis P.H., 2008, in {\it Dynamics and Thermodynamics of systems with long-range interactions: Theory and Experiments}, AIP Conf. Proc., 970, 39

\bibitem[\protect\citeauthoryear{Ciotti} {1991}]{ciotti91} Ciotti L., 1991, A\&A, 249, 99

\bibitem[\protect\citeauthoryear{Ciotti} {2000}]{ciotti2000} Ciotti L., 2000, Lecture Notes on Stellar Dynamics. Scuola Normale Superiore, Pisa

\bibitem[\protect\citeauthoryear{Ciotti} {2009}]{ciotti09} Ciotti L., 2009, NCimR, 32, 1

\bibitem[\protect\citeauthoryear{Ciotti \& Bertin} {1999}] {ciotti99} Ciotti L. \& Bertin, G., 1999, A\&A, 352, 447

\bibitem[\protect\citeauthoryear{Ciotti {\rm et al.}} {2006}]{Ciotti06} Ciotti L.,  Londrillo, P. \& Nipoti, C.,  2006, ApJ, 640, 741.

\bibitem[\protect\citeauthoryear{Ciotti \& Morganti} {2010a}]{CM10a} Ciotti L. \& Morganti L.,  2010a, MNRAS, 401, 1091.

\bibitem[\protect\citeauthoryear{Ciotti \& Morganti} {2010b}]{CM10b} Ciotti L. \& Morganti L.,  2010b, MNRAS, 408, 1070.

\bibitem[\protect\citeauthoryear{Ciotti {\rm et al.}} {2007}]{CNL07} Ciotti L., Nipoti C., Londrillo P.,  2007, {\it Proc. Int. Workshop on Collective phenomena in macroscopic systems}, World Scientific 177.

\bibitem[\protect\citeauthoryear{Courteau et al.} {1996}]{courteau96} Courteau S., de Jong R.S., Broeils A.H., 1996, ApJ, 437, 21

\bibitem[\protect\citeauthoryear{Dehnen} {2001}]{DEH01} Dehnen W., 2001, MNRAS, 324, 273

\bibitem[\protect\citeauthoryear{Di Cintio} {2009}]{dicintio2009} Di Cintio P.F., 2009, Master Thesis, Bologna University (DC09)

\bibitem[\protect\citeauthoryear{Di Cintio \& Ciotti} {2011}]{dicintio2011} Di Cintio P.F. \& Ciotti L., 2011, IJBC, 21, 2279 (DCC11)

\bibitem[\protect\citeauthoryear{Gabrielli et al.} {2010}]{gabri10} Gabrielli A., Joyce M., Marcos B., 2010, Ph.Rev.Lett., 105, 210602

\bibitem[\protect\citeauthoryear{Gerhard \& Spergel} {1992}]{spergel} Gerhard O.E. \& Spergel D.N., 1992, ApJ, 397, 38

\bibitem[\protect\citeauthoryear{Graham \& Colless} {1997}]{graham97} Graham A. \& Colless M.,  1997, MNRAS, 287, 221

\bibitem[\protect\citeauthoryear{Graham} {1998}]{graham98} Graham A., 1998, MNRAS, 293, 933

\bibitem[\protect\citeauthoryear{H\'enon} {1964}]{hen64} H\'enon M., 1964, Ann. d'Astroph., 27, 83

\bibitem[\protect\citeauthoryear{Hernquist} {1990}]{her} Hernquist L., 1990, ApJ, 356, 359

\bibitem[\protect\citeauthoryear{Hansen \& Moore} {2006}]{hanmoo} Hansen S.H. \& Moore, B. 2006, New Astronomy, 11, 333

\bibitem[\protect\citeauthoryear{Hansen {\rm et al.}} {2010}]{han} Hansen S.H.,  Juncher D., Sparre M., 2010, ApJ, 718, 68

\bibitem[\protect\citeauthoryear{Joyce et al.} {2009}]{joyce09} Joyce M., Marcos B., Sylos Labini F., 2009, MNRAS, 397, 775

\bibitem[\protect\citeauthoryear{Joyce \& Worrakitpoonpon} {2011}]{joyce10} Joyce M. \& Worrakitpoonpon T., 2011, Ph.Rev.E., 84, 1139

\bibitem[\protect\citeauthoryear{Kandrup} {1989}]{kan89} Kandrup H.E., 1989, Ph.Rev.A., 40, 7265

\bibitem[\protect\citeauthoryear{Lynden-Bell} {1967}]{LB67} Lynden-Bell D., 1967, MNRAS, 136, 101

\bibitem[\protect\citeauthoryear{Lynden-Bell \& Lynden-Bell} {1982}]{LB82} Lynden-Bell D. \& Lynden-Bell R. M., 1982, Proceedings: Mathematical, Physical and Engineering Sciences, Vol. 455, No. 1982, p. 475, The Royal Society

\bibitem[\protect\citeauthoryear{Londrillo {\rm et al.}} {1991}]{LMS91} Londrillo P., Messina A.,  Stiavelli, M 1991, MNRAS, 250, 54

\bibitem[\protect\citeauthoryear{Londrillo {\rm et al.}} {2003}]{fvfps} Londrillo P., Nipoti C., Ciotti L., 2003, MSAIS, 1, 18

\bibitem[\protect\citeauthoryear{Londrillo \& Nipoti} {2009}]{NMODY} Londrillo P. \& Nipoti C., 2009, MSAIS, 13, 89

\bibitem[\protect\citeauthoryear{Ludlow {\rm et al.}} {2010}]{Lud10}  Ludlow A.D., Navarro J.F., Springel V., Vogelsberger M., Wang J., White S.D.M., Jenkins A., Frenk C.S.,  2010, MNRAS, 406, 137

\bibitem[\protect\citeauthoryear{Malekjani {\rm et al.}} {2009}]{MRH09} Malekjani M., Rahvar S., Haghi H., 2009, ApJ, 694, 1220

\bibitem[\protect\citeauthoryear{Malekjani {\rm et al.}} {2012}]{MHJ12} Malekjani M., Haghi H., Jassur D.M.Z., 2012, New Astronomy, 17, 149

\bibitem[\protect\citeauthoryear{Marcos et al.} {2012}]{gabri12} Marcos B., Gabrielli A., Joyce M.,  2012, CEJPh, 10, 676

\bibitem[\protect\citeauthoryear{Meza \& Zamorano} {1997}]{MZ97} Meza A., \& Zamorano N., 1997, ApJ, 490, 136

\bibitem[\protect\citeauthoryear{Milgrom} {2010}]{Mil10} Milgrom M.,  2010, MNRAS, 403, 886

\bibitem[\protect\citeauthoryear{Mineau {\rm et al.}} {1990}]{mineau90} Mineau P., Feix M.R., Rouet, J.L., 1990, A\&A, 228, 344

\bibitem[\protect\citeauthoryear{Moffat \& Sokolov} {1996}]{moff96} Moffat J.W. \& Sokolov I.Yu., 1996, Phys.Lett.B, 378, 59

\bibitem[\protect\citeauthoryear{Navarro {\rm et al.}} {1997}]{NFW} Navarro J.F., Frenk C.S., White S.D.M., 1997, ApJ, 490, 493

\bibitem[\protect\citeauthoryear{Nipoti {\rm et al.}} {2002}]{NLC02} Nipoti C., Londrillo  P., Ciotti L., 2002, MNRAS, 332, 901

\bibitem[\protect\citeauthoryear{Nipoti {\rm et al.}} {2006a}]{NLC06a} Nipoti C., Londrillo  P., Ciotti, L., 2006, MNRAS, 370, 681 (N06a)

\bibitem[\protect\citeauthoryear{Nipoti {\rm et al.}} {2006b}]{NLC06b} Nipoti, C., Londrillo,  P. \& Ciotti, L. 2006, {\it Science and Supercomputing at CINECA}, 122 (N06b)

\bibitem[\protect\citeauthoryear{Nipoti {\rm et al.}} {2007a}]{NLC07} Nipoti C., Londrillo P., Ciotti L., 2007a, ApJ, 660, 256 (N07a)

\bibitem[\protect\citeauthoryear{Nipoti {\rm et al.}} {2007b}]{NLC07bis} Nipoti C., Londrillo P., Ciotti L., 2007b, MNRAS, 381, 107

\bibitem[\protect\citeauthoryear{Nipoti {\rm et al.}} {2011}]{NCL11} Nipoti C., Ciotti L., Londrillo  P.,  2011, MNRAS, 414, 3298

\bibitem[\protect\citeauthoryear{Plummer} {1911}]{Plu} Plummer H. L., 1911, MNRAS, 71, 460

\bibitem[\protect\citeauthoryear{Prugniel \& Simien} {1997}]{prugniel97}  Prugniel P. \& Simien F., 1997, A\&A, 321, 111

\bibitem[\protect\citeauthoryear{Sanders} {1998}]{Sanders98} Sanders R.H., 1998, MNRAS, 296, 1009

\bibitem[\protect\citeauthoryear{Sanders} {2008}]{Sanders08} Sanders R.H., 2008, MNRAS, 386, 1588

\bibitem[\protect\citeauthoryear{Sparre \& Hansen} {2012}]{sparre12} Sparre M. \& Hansen, S.H., 2012, JCAP, 10, 49

\bibitem[\protect\citeauthoryear{Srinirasan et al.} {2005}]{srinirasan} Srinivasan K., Mahawar H., Sarin V., 2005, ICCS, LNCS3514, p.107-114,  V.S. Sunderam et al. (Eds.) Springer-Verlag Berlin

\bibitem[\protect\citeauthoryear{Stein} {1970}]{stein70} Stein E.M., 1970, Singular integrals and differentiability properties of functions, (Princeton University Press)

\bibitem[\protect\citeauthoryear{Sylos Labini} {2013}]{fsl13} Sylos Labini F.,  2013, MNRAS, 429, 679

\bibitem[\protect\citeauthoryear{Takizawa \& Inagaki} {1997}]{tak97} Takizawa M. \& Inagaki S., 1997, arXiv:astro-ph/9702002v1

\bibitem[\protect\citeauthoryear{Taylor \& Navarro} {2001}]{tana01} Taylor J.E. \& Navarro J.F., 2001, ApJ, 563, 483

\bibitem[\protect\citeauthoryear{Teles {\rm et al.}} {2011}]{TELE11} Teles T.N., Levin Y.,  Pakter, R.,   2011,  MNRAS, 417, 21

\bibitem[\protect\citeauthoryear{Trenti \& Bertin} {2005}]{bt05} Trenti M. \& Bertin G., 2005, A\&A, 429, 161

\bibitem[\protect\citeauthoryear{Trenti {\rm et al.}} {2005}]{bert05} Trenti M., Bertin G., van Albada T.S., 2005, A\&A 433, 57

\bibitem[\protect\citeauthoryear{Trujillo {\rm et al.}} {2001}]{trujillo01} Trujillo I., Graham, A., Caon M., 2001, MNRAS, 326, 869

\bibitem[\protect\citeauthoryear{van Albada} {1982}]{vanAlbada82} van Albada T.S., 1982, MNRAS, 201, 939

\bibitem[\protect\citeauthoryear{van Hese {\rm et al.}} {2011}]{vanhese11} van Hese E., Baes M., Dejonghe H., 2011, ApJ, 726, 80

\bibitem[\protect\citeauthoryear{Visbal et al.} {2012}]{visbal} Visbal E., Loeb A., Hernquist L., 2012, arXiv:astro-ph/1206.5852

\bibitem[\protect\citeauthoryear{Youngkins \& Miller} {2000}]{YM2000} Youngkins V.P. \& Miller B.N., 2000, Ph.Rev.E, 62, 4583 

\bibitem[\protect\citeauthoryear{Zocchi} {2010}]{zocchi} Zocchi A., 2010, Master Thesis, Milano University

\end{thebibliography}
\end{document}